\documentclass{PoS}
\usepackage[caption=false]{subfig}
\usepackage{graphicx}                                
\usepackage{amsmath,amsfonts,bbm}
\usepackage{amssymb}
\usepackage{mathrsfs}

\newcommand{\pslash}{p\hspace{-2mm}\slash}
\newcommand{\vecpslash}{\vec p\hspace{-2mm}\slash}
\newcommand{\intp}{\int dp_4\,}
     
\title{Confinement in Coulomb gauge}

\ShortTitle{Confinement in Coulomb gauge}

\author{\speaker{G.~Burgio}\\
        Institut f\"ur Theoretische Physik\\
        Auf der Morgenstelle 14\\
        72076 T\"ubingen\\
        Germany\\
        E-mail: \email{giuseppe.burgio@uni-tuebingen.de}}
\author{Markus Quandt\\
        Institut f\"ur Theoretische Physik\\
        Auf der Morgenstelle 14\\
        72076 T\"ubingen\\
        Germany\\
        E-mail: \email{markus.quandt@uni-tuebingen.de}}
\author{Hugo Reinhardt\\
        Institut f\"ur Theoretische Physik\\
        Auf der Morgenstelle 14\\
        72076 T\"ubingen\\
        Germany\\
        E-mail: \email{hugo.reinhardt@uni-tuebingen.de}}
\author{Mario Schr\"ock\\
        Institut f\"ur Physik, FB Theoretische Physik\\
        Universit\"at Graz\\
        8010 Graz\\
        Austria\\
        E-mail: \email{mario.schroeck@uni-graz.at}}
\author{H.~Vogt\\
        Institut f\"ur Theoretische Physik\\
        Auf der Morgenstelle 14\\
        72076 T\"ubingen\\
        Germany\\
        E-mail: \email{hannes.vogt@uni-tuebingen.de}}

\abstract{We review our lattice results concerning the Gribov-Zwanziger 
confinement mechanism in Coulomb gauge. In particular, we verify the validity 
of Gribov's IR divergence condition for the Coulomb ghost form factor. We also 
show how the quark self-energy is, like that of the transverse gluon, IR 
divergent, thus effectively extending the Gribov-Zwanziger scenario to full 
QCD.}

\FullConference{31st International Symposium on Lattice Field Theory - 
LATTICE 2013\\
July 29 - August 3, 2013\\
Mainz, Germany}

\begin{document}

\section{Introduction}

As Gribov first to noticed \cite{Gribov:1977wm}, 
for non-Abelian theories most gauge conditions admit several solutions 
and the corresponding Faddeev-Popov 
(FP) mechanism is not sufficient to define the functional integral 
beyond perturbation theory. The field-configuration space must therefore be 
restricted to a domain, continuously connected to the origin, where the gauge 
condition possesses unique solutions. He then showed how, as 
soon as the fields cross the boundary of such region, the ghost dressing 
function acquires a singularity; his ``no-pole'' condition for the FP-ghost 
at non-vanishing momentum is then necessary to implement the restriction
to the so called Gribov region. 

In particular, in Coulomb 
gauge, he argued how such restriction can imply a diverging gluon self-energy, 
motivating its disappearance from the physical spectrum. 
QCD in Coulomb gauge has since then been a fruitful playground to 
investigate the Gribov-Zwanziger (GZ) confinement mechanism
\cite{Gribov:1977wm,Zwanziger:1995cv}. In a series of papers 
\cite{Burgio:2008jr,Burgio:2009xp,Quandt:2010yq,Reinhardt:2011fq,%
Burgio:2012ph,Burgio:2012bk}, briefly summarized here, we have 
analyzed the behaviour of the relevant two-point functions 
at zero temperature on the lattice and 
compared them with the corresponding predictions of Hamiltonian variational 
calculations \cite{Szczepaniak:2001rg,Feuchter:2004mk,Epple:2006hv}.

Gribov based his 
conjectures on more or less heuristic arguments, which Zwanziger later tried to 
put on a more solid basis, while variational calculations, 
viable in Coulomb gauge since they by-pass the explicit
construction of the gauge invariant Hilbert space \cite{Burgio:1999tg}, did 
provide some insight on the relation of the GZ-mechanism to the Hamiltonian 
formulation. In both cases, however, approximations need to be made; although 
many authors tackled the problems during the years \cite{Cucchieri:2000kw,%
Langfeld:2004qs,Nakagawa:2007fa,Voigt:2008rr,Nakagawa:2008zza,%
Nakagawa:2008zzc}, a satisfactory non-perturbative cross-check from lattice 
calculation was hindered for a long time by the presence of strong 
discretization effects. We have shown \cite{Burgio:2008jr,%
Burgio:2012ph,Burgio:2012bk} how for each propagator a mixture of improved 
actions and separate treatment of their energy dependence can quite 
effectively solve such problems, allowing an explicit check of the GZ-scenario.
In particular, anisotropic actions prove to be very useful
\cite{Burgio:1996ji,Burgio:2003in}; details can be
found in \cite{Burgio:2012bk}, as well as a description of the gauge fixing
algorithm, which adapts those introduced in 
\cite{Bogolubsky:2005wf,Bogolubsky:2007bw}.
Following the ideas proposed in \cite{Burgio:2008jr}, a first 
analysis of the SU(3) case had been attempted in 
\cite{Nakagawa:2011ar}.

From the continuum analysis and from Ref.~ \cite{Burgio:2008jr,Burgio:2009xp} 
we know that in the pure gauge
sector the static gluon propagator, the static Coulomb potential and the 
ghost form factor should obey:
\begin{equation}
\begin{array}{l}
D(\vec{p})\; \simeq \frac{\displaystyle |\vec{p}|}{\displaystyle 
\sqrt{|\vec{p}|^4+M^4}}\\
\\
V_C(\vec{p})  \simeq \frac{\displaystyle 8 \pi \sigma_C}{\displaystyle |{\vec{p}}|^4}
 + \frac{\displaystyle \eta}{\displaystyle |{\vec{p}}|^2}
      + \mathcal{O}(1)
\end{array}
\qquad
d(\vec{p}) \simeq \left\{\begin{array}{c l}
\frac{\displaystyle 1}{\displaystyle |\vec{p}|^{\kappa_{\rm gh}}}&\quad|\vec{p}|\ll \Lambda\\ 
&\\
\frac{\displaystyle 1}{\displaystyle \log^{\gamma_{\rm gh}}{{|\vec{p}|}}} &\quad|\vec{p}|\gg \Lambda
\end{array}\right.
\label{eq_1_1}
\end{equation}
where the Gribov mass $M\simeq 1$~GeV; the gluon self-energy is given by 
$\omega_A = D^{-1}$. 
The quark propagator, the fermion self-energy and the 
running mass $M(|\vec{p}|)$ will take the form \cite{Burgio:2012ph}:
\begin{equation}
\begin{array}{c}
 S(\vec{p},p_4) = \frac{\displaystyle Z(\vec{p})}{\displaystyle i\vecpslash 
      + i\pslash_4 \alpha(\vec{p})+M(\vec{p})}\qquad
\omega_F(|\vec{p}|) =
    \frac{\displaystyle \alpha(|\vec{p}|)}{\displaystyle Z^n(|\vec{p}|)}\sqrt{\vec{p}^2 + 
      M^2(|\vec{p}|)}\\
\\
M(|\vec{p}|) = \frac{\displaystyle m_\chi(m_b)}{\displaystyle 1+b\, \frac{|\vec{p}|^2}{\Lambda^2}\,
\log{\left(e+\frac{|\vec{p}|^2}{\Lambda^2}\right)}^{-\gamma}}+
\frac{\displaystyle m_r(m_b)}{\displaystyle \log{\left(e+\frac{|\vec{p}|^2}{\Lambda^2}\right)}^{\gamma}}\,,
\end{array}
\label{eq_1_2}
\end{equation}
where $Z$ is the field renormalization function, $\alpha$ the energy
renormalization function, $m_b$ the bare quark mass, $m_\chi(m_b)$ the chiral mass
and $ m_r(m_b)$ the renormalized current mass \cite{Burgio:2012ph}. The exponent
$n$ in the rhs of $\omega_F$ depends on the exact definitions of the
static quark propagator and its self-energy. In the Hamitonian 
quasi-particle approach,
with which we ultimatively wish to compare, the fermion 
propagator, taken as an operator, is proportional to the Hamiltonian, where 
the inverse of the proportionality coefficient gives the eigenvalue of $H$, 
i.e. the
quark effective energy:
\begin{eqnarray}
S^H(\vec{p}) &=& \intp S(\vec{p},p_4) = 
\frac{\displaystyle Z(\vec{p})}{\displaystyle 
        \alpha(\displaystyle \vec{p})} \frac{\displaystyle \sqrt{\vec{p}^2 
          + M^2(|\vec{p}|)}}{\displaystyle i\vecpslash + M(\vec{p})}= 
      \frac{\displaystyle Z(\vec{p})}{\displaystyle 
        \alpha(\displaystyle \vec{p})} \frac{\displaystyle -i\vecpslash +
          M(\vec{p})}{\displaystyle \sqrt{\vec{p}^2 + M^2(|\vec{p}|)}}\\
        \omega^H_\psi(|\vec{p}|)&=& \frac{\alpha(|\vec{p}|)}{Z(|\vec{p}|)}
      \sqrt{\vec{p}^2 + M^2(|\vec{p}|)}\,.
\end{eqnarray}
Alternatively, in analogy with the euclidean free fermion case, one can invoke
consistency between $\intp S(\vec{p},p_4)$ and $\intp S^{-1}(\vec{p},p_4)$,
leading to 
\begin{equation}
S^E(\vec{p}) = \Lambda \frac{\displaystyle Z(\vec{p})}{\displaystyle i\vecpslash +M(\vec{p})}
\end{equation}
where however $\Lambda \propto a^{-1}_t$, so that an extra renormalization
is needed. In this case the 
quark effective energy would be given by:
\begin{equation}
\omega^E_\psi(|\vec{p}|) = \intp S^2(\vec{p},p_4)=
\frac{\alpha(|\vec{p}|)}{Z^2(|\vec{p}|)}\sqrt{\vec{p}^2 + 
M^2(|\vec{p}|)}\,.
\end{equation}
Eq.~(\ref{eq_1_2}) summarizes the two alternative definitions.

\section{Results}

\subsection{Ghost form factor}

\begin{figure}[htb]
\subfloat[][]{\includegraphics[width=0.49\textwidth,height=0.44\textwidth]{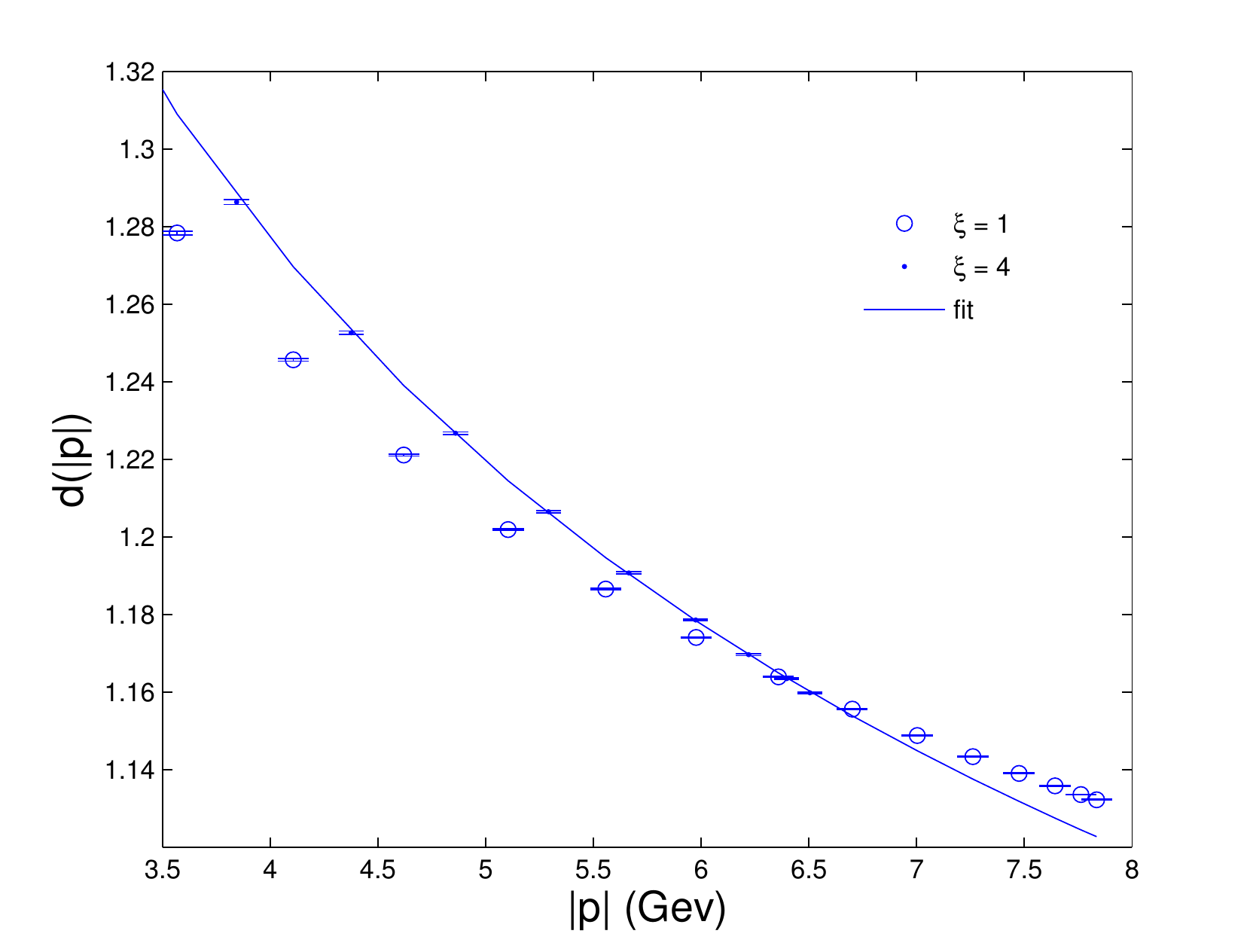}}
\subfloat[][]{\includegraphics[width=0.49\textwidth,height=0.44\textwidth]{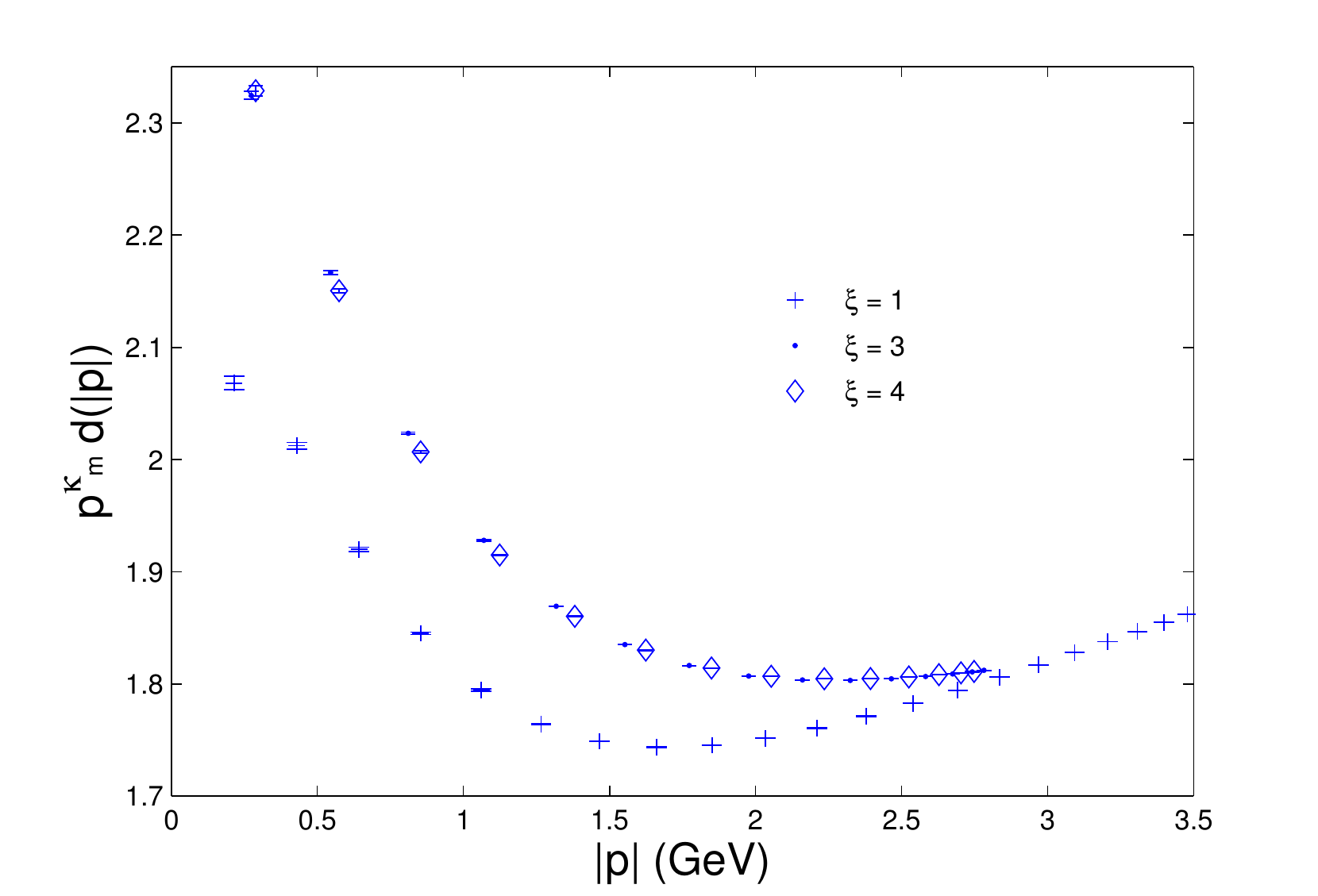}}
\caption{(a): UV behavior of $ d({\vec{p}})$ compared with 
Eq.~(\protect\ref{eq_1_1}). (b): IR behavior of 
$|{\vec{p}}|^{\kappa_m}\, d({\vec{p}})$, both for different anisotropies $\xi$.}
\label{fig1}
\end{figure}

A careful analysis of the ghost form factor in the Hamiltonian limit 
$a_t \to 0$ 
shows that its UV behaviour agrees with Eq.~(\ref{eq_1_1}), with 
$\gamma_{\rm gh} =1/2$, confirming continuum predictions, and $m = 0.21(1)$~GeV, 
see Fig.~\ref{fig1}~(a). In the IR going to higher anisotropies increases
the exponent $\kappa_{\rm gh}$, as shown in Fig.~\ref{fig1}~(b), where we plot
$|\vec{p}|^{\kappa_m} \, d(\vec{p})$, with $\kappa_m$ the IR exponent
for $\xi=1$, as a function 
of the anisotropy. The limit $\xi\to\infty$ gives
$\kappa_{\rm gh} \gtrsim 0.5$, confirming the GZ-scenario. This however 
disagrees with some continuum predictions $\kappa_{\rm gh} =1$, deriving from 
the assumption of the finiteness of the static ghost-gluon vertex. Whether 
this is indeed correct and algorithmic improvements could change the lattice 
result is still a matter of investigation, see H.~Vogt's talk \cite{Vogt:2013jha}.

\subsection{Coulomb potential}

\begin{figure}[htb]
\subfloat[][]{\includegraphics[width=0.49\textwidth,height=0.44\textwidth]{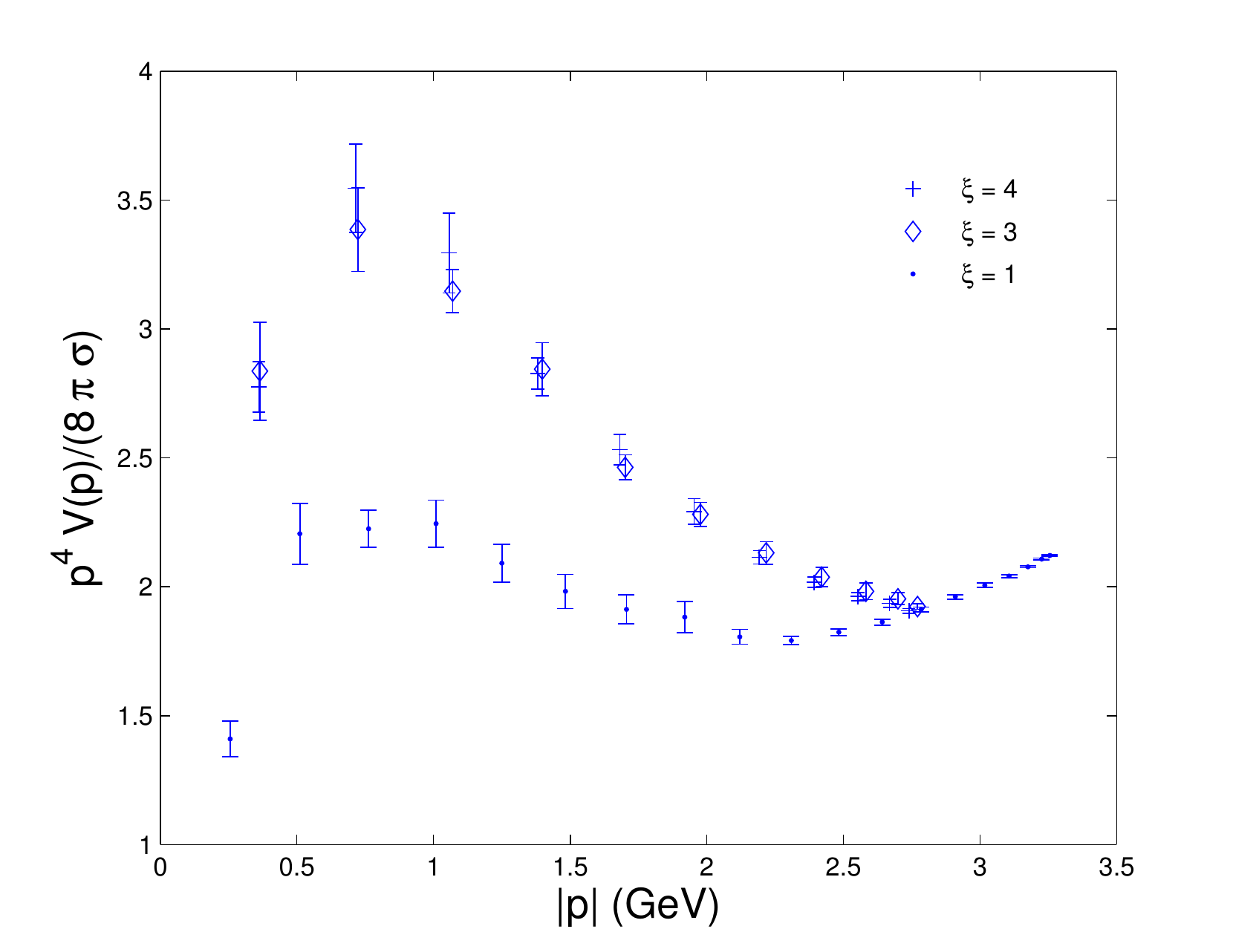}}
\subfloat[][]{\includegraphics[width=0.48\textwidth,height=0.43\textwidth]{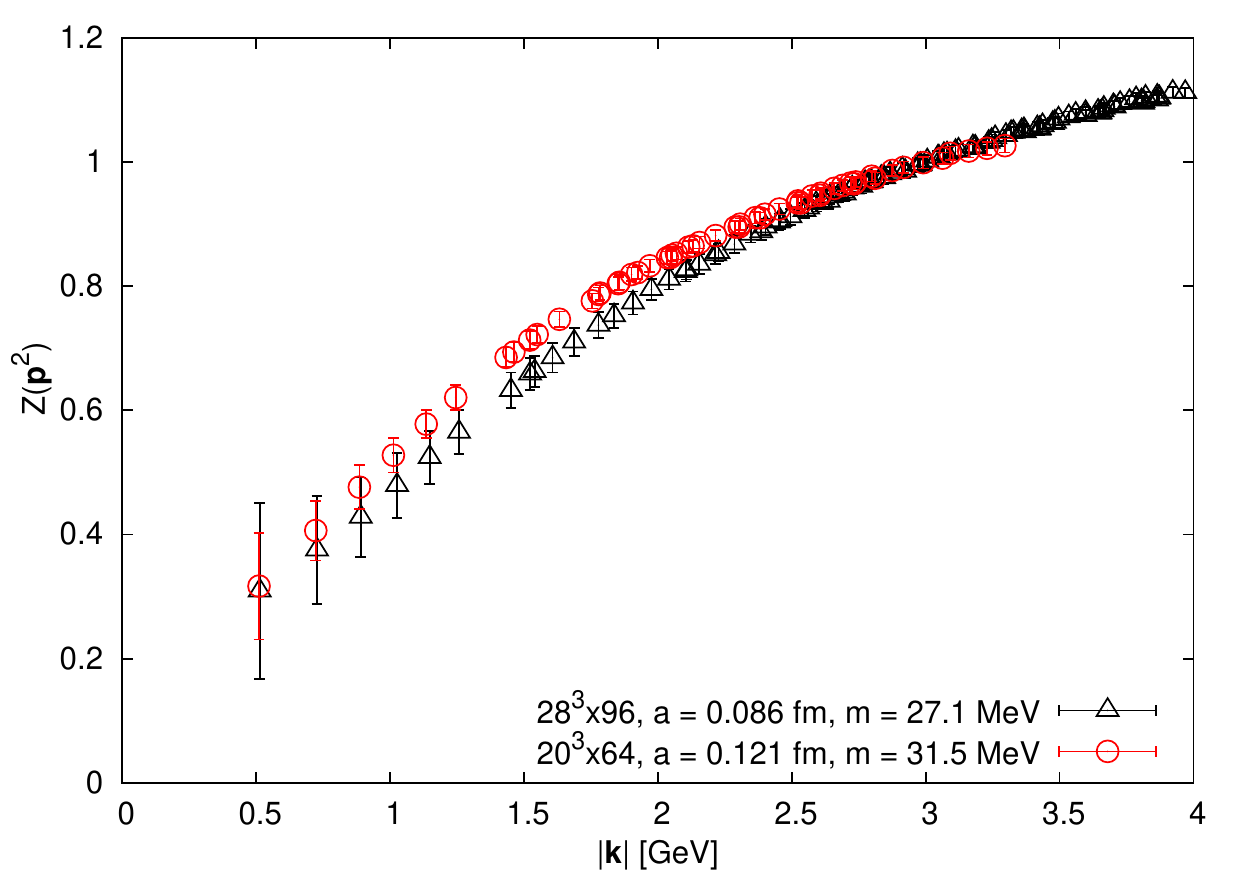}}
\caption{{(a)}: Infrared behavior of $|{\vec{p}}|^4\, V_C(\vec{p})/(8 \pi \sigma)$ for 
different anisotropies $\xi$. {(b)}: Quark field renormalization
function $Z(|{\vec{p}}|)$.}
\label{fig3}
\end{figure}

In Fig.~\ref{fig3}~(a) we show $|{\vec{p}}|^4 V_C(|{\vec{p}}|)$ as obtained 
from different anisotropies. Somewhat boldly fitting the results to 
Eq.~(\ref{eq_1_1}) we get,
in the Hamiltonian limit $\xi\to\infty$, $\sigma_C = 2.2(2) \,\sigma$ as
expected from Zwanziger's predictions \cite{Zwanziger:2002sh}; see H.~Vogt's 
talk \cite{Vogt:2013jha} in this conference for a more ``honest'' discussion
of the difficulties involved.

\subsection{Quark propagator}

Our calculations were all made on a set of configurations generated by the MILC
collaboration \cite{Bazavov:2009bb}, see \cite{Burgio:2012ph} for details.
The use of improved actions is crucial to establish the scaling
properties of the Coulomb gauge quark propagators. This is very
similar to the situation in Landau gauge, see e.g. 
\cite{Bowman:2002bm,Bowman:2005vx,Parappilly:2005ei}, whose techniques
we have adapted to our case. 

\begin{figure}[htb]
\subfloat[][]{\includegraphics[width=0.49\textwidth,height=0.44\textwidth]{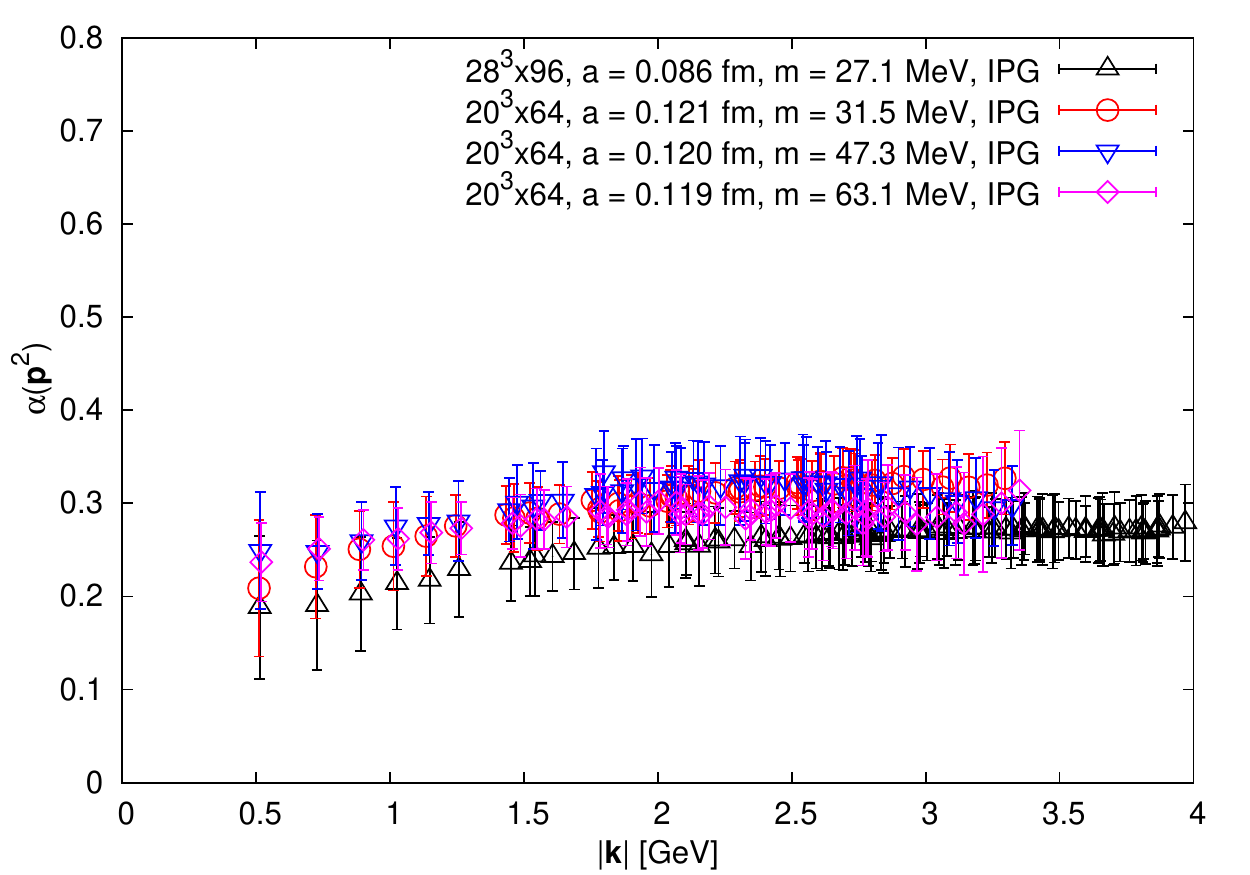}}
\subfloat[][]{\includegraphics[width=0.49\textwidth,height=0.44\textwidth]{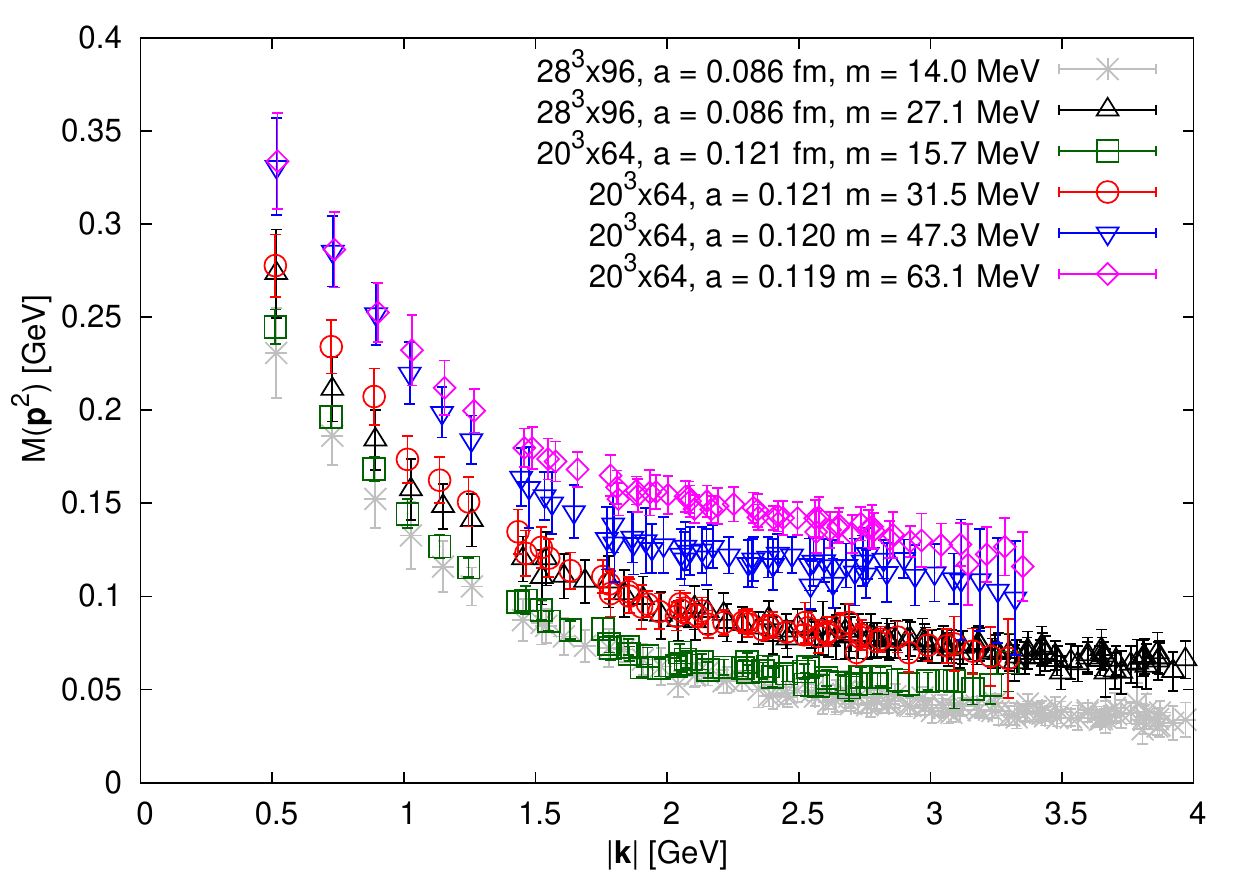}}
\caption{{(a)}: Energy renormalization function $\alpha(|{\vec{p}}|)$. {(b)}: 
Running mass $M(|{\vec{p}}|)$.}
\label{fig4}
\end{figure}

Fig.~\ref{fig3}~(b) shows the scaling of the renormalization function 
$Z(|{\vec{p}}|)$ for configurations calculated at similar bare quark mass, 
while the RG-invariant functions $\alpha(|{\vec{p}}|)$ and $M(|{\vec{p}}|)$
are given in Fig.~\ref{fig4}. Their behaviour agrees with theoretical
expectations, see Eq.~(\ref{eq_1_2}).

\begin{figure}[htb]
\subfloat[][]{\includegraphics[width=0.49\textwidth,height=0.44\textwidth]{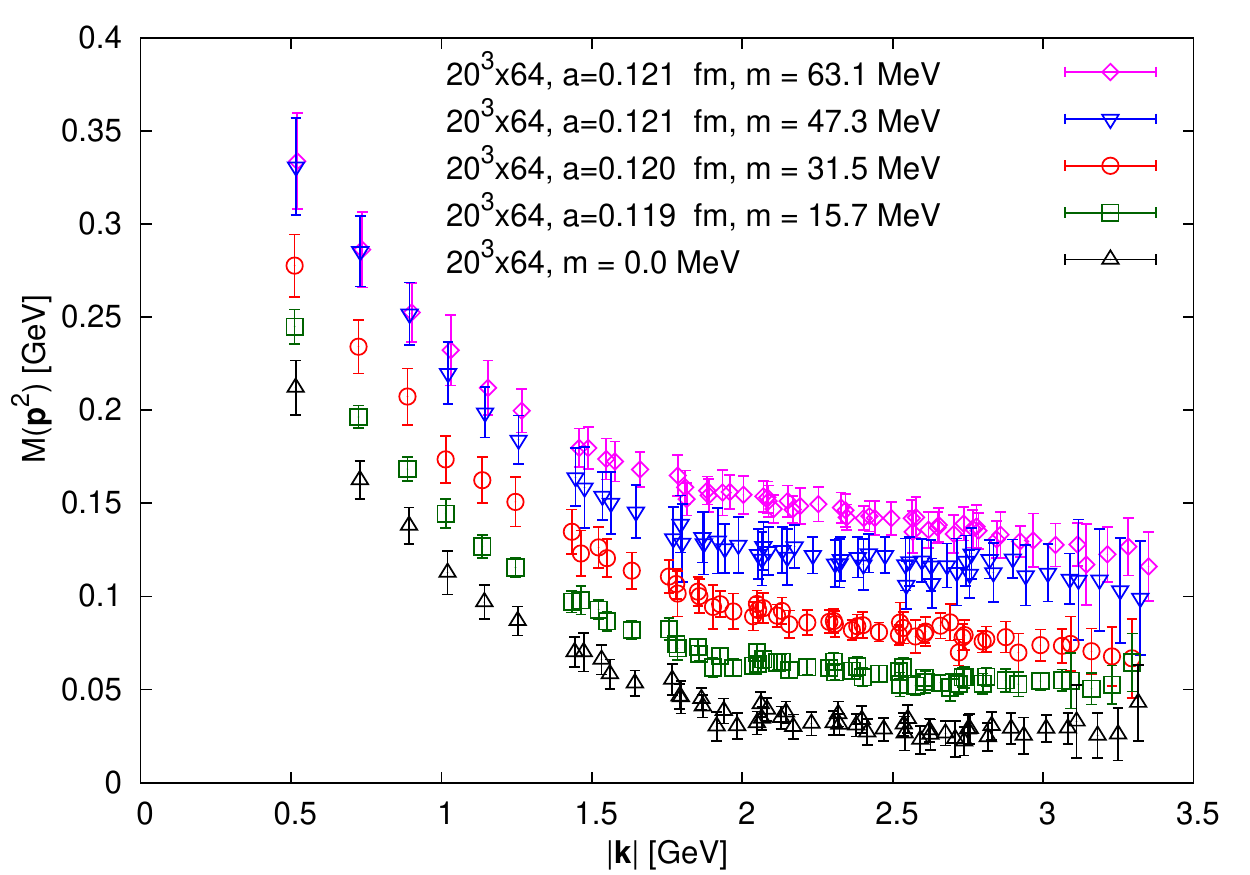}}
\subfloat[][]{\includegraphics[width=0.49\textwidth,height=0.44\textwidth]{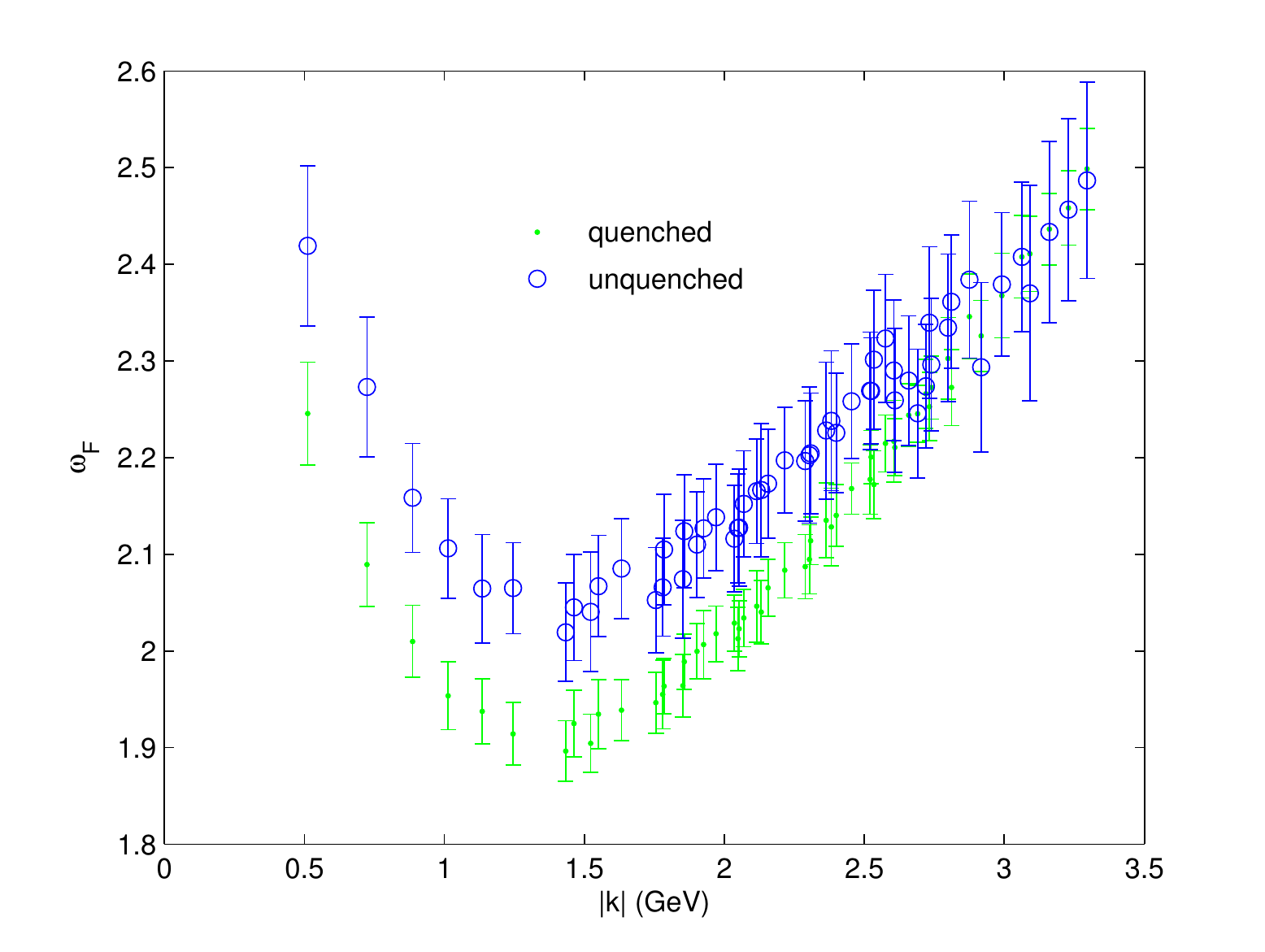}}
\caption{{(a)}: Quark self-energy $\omega_F(|\vec{p}|)$. {(b)}: Running
mass $M(|{\vec{p}}|)$ in the chiral limit $m_b\to 0$; see
Eq.~(\protect\ref{eq_1_2}).}
\label{fig5}
\end{figure}

Our most interesting results are given in Fig.~\ref{fig5}. Analogously to
the gluon self-energy $\omega_A(|\vec{p}|)$, the quark self-energy 
$\omega_F(|\vec{p}|)$ has a turn-over at $|\vec{p}| \sim 1$~GeV, 
clearly departing from the behaviour of a free particle, and diverging in the 
IR, see Fig.~\ref{fig5}~(a). Such results does not depend on the exact 
definition of $\omega_F$, since it only relies on the IR vanishing of $Z(p)$,
see Fig.~\ref{fig3}~(b).
Awaiting confirmation on larger 
lattices, this would extend the Gribov argument to full QCD. Moreover, as 
Fig.~\ref{fig5}~(b) shows, the running mass $M(|{\vec{p}}|)$ is 
quantitatively compatible with our phenomenological expectations
from chiral symmetry breaking. Fitting it to Eq.~(\ref{eq_1_2})
we obtain $b = 2.9(1)$, $\gamma = 0.84(2)$, $\Lambda = 1.22(6)$~GeV, 
$m_\chi(0) = 0.31(1)$~GeV, with $\chi^2$/d.o.f.$=1.06$. 

\section{Conclusions}

We have shown that lattice calculations confirm the GZ confinement scenario 
in Coulomb gauge at $T=0$. The ghost form factor $d(|\vec{p}|)$ is IR divergent
with an exponent $\kappa_{\rm gh} \gtrsim 0.5$, implying Gribov's no-pole 
condition and a
dual-superconducting scenario \cite{Reinhardt:2008ek}; 
the gluon propagator satisfies
the Gribov formula, implying an IR diverging self-energy, and the Coulomb 
potential seems compatible with a string tension roughly twice the physical 
string tension. Moreover from the 
quark propagator we can extract the quark self-energy 
$\omega_F(|\vec{p}|)$, which is compatible with an IR divergent behaviour, 
and the running mass $M(|{\vec{p}}|)$, which gives a constituent quark mass
of $m_\chi(0) = 0.31(1)$~GeV.

The situation in Coulomb gauge seem to be easier than in Landau gauge, 
where BRST symmetry is non-perturbatively broken, violating the Kugo-Ojima
confinement scenario \cite{Kugo:1979gm}, while the GZ confinement scenario 
is realized explicitly introducing of an horizon function, 
see e.g. \cite{Vandersickel:2012tz} for a recent review; its physical
implications and how these can be related to the presence of
dim-2 condensates \cite{Burgio:1997hc,Akhoury:1997by,Boucaud:2000ey}
are an interesting issue still debated in the literature 
\cite{Cucchieri:2011ig}.

Of course, an extension of the above results to finite temperature is the
``missing link'' to a full understanding of confinement in Coulomb gauge. 
First attempts in this direction and the related difficulties are discussed
in H.~Vogt's talk \cite{Vogt:2013jha}.

\section*{Aknowledgements}
This work was partially supported the DFG under the contract DFG-Re856/6-3.

\end{document}